\documentclass[oneside,reqno,12pt]{amsart}

\usepackage{rangecite}
\usepackage{euscript}

\newcommand{\mathscript}{\EuScript}

\numberwithin{equation}{section}

\newcommand{\bg}{\bar{g}}
\newcommand{\bR}{\bar{R}}
\newcommand{\bn}{\bar{\nabla}}
\newcommand{\e}{\epsilon}

\newcommand{\D}{\mathscript{D}}
\newcommand{\Lag}{\mathscript{L}}
\newcommand{\diff}[2]{\frac{\partial #1}{\partial #2}}

\newcommand{\du}[3]{{\/{#1}_{#2}}^{#3}}
\newcommand{\ud}[3]{{\/{#1}^{#2}}_{#3}}
\newcommand{\udu}[4]{{\smash{{{#1}^{#2}}_{#3}}}^{#4}}

\makeatletter
    \def\serieslogo@{\vtop to 0pt{\noindent\scriptsize\ppn\parindent\z@}}
    \let\@setcopyright\@empty
\makeatother

\begin{document}

\def\ppn{HEP-TH/9509174, UPR-663T}

\vspace*{0.2in}

\title[SUSY Breaking in Two-Dimensional Supergravity]{Two-Dimensional 
Higher-Derivative Supergravity and \\ a New Mechanism for Supersymmetry 
Breaking}
\author{Ahmed Hindawi, Burt A. Ovrut, and Daniel Waldram}
\thanks{Published in Nuclear Physics B \textbf{471} (1996), 409--429}
\maketitle
\vspace*{-0.3in}
\begin{center}
\small{\textit{Department of Physics, University of Pennsylvania}} \\
\small{\textit{Philadelphia, PA 19104-6396, USA}}
\end{center}
\begin{abstract}

We discuss the general form of quadratic $(1,1)$ supergravity in two 
dimensions, and show that this theory is equivalent to two scalar 
supermultiplets coupled to non-trivial supergravity. It is demonstrated 
that the theory possesses stable vacua with vanishing cosmological 
constant which spontaneously break supersymmetry.

\vspace*{\baselineskip}



\end{abstract}

\renewcommand{\baselinestretch}{1.2} \large \normalsize

\vspace*{\baselineskip}

\section{Introduction}

In two recent papers \cite{PRD-53-5583,PRD-53-5597}, we discussed theories of 
higher-derivative bosonic gravitation in four dimensions. Such theories 
have higher-order equations of motion for the metric and describe, in 
addition to the helicity-two graviton, extra scalar and symmetric-tensor 
degrees of freedom. We presented a method for reducing such theories to 
a canonical second-order form by introducing the new degrees of freedom 
explicitly through a Legendre transformation, the exact analog of 
forming the Helmholtz Lagrangian to reduce a second-order theory to 
first-order. Using the second-order form, we explored the vacuum 
structure of these theories and showed, in particular, the existence of  
non-trivial vacua which have a non-negligible effect on low-energy 
physics. However, it turned out that all such non-trivial vacua must 
have non-vanishing cosmological constant and, hence, correspond to 
either deSitter or anti-deSitter spacetime with a radius generically of 
the order of the inverse Planck mass. It is little wonder that such 
non-trivial gravity vacua have played no role in particle physics to 
date.

In this paper, we continue to explore higher-derivative gravitational 
theories but with two modifications. First, we restrict our discussion 
to two dimensions and second, and most importantly, we introduce 
supersymmetry, analyzing the vacuum structure of higher-derivative 
$(1,1)$ supergravity. We find that these theories continue to exhibit 
non-trivial vacua, with a richer structure, in fact, than in the bosonic 
case. Remarkably, we find that these non-trivial vacua can now have 
vanishing cosmological constant and, hence, correspond to flat 
spacetime. Exactly what this means, and how it occurs, will be 
explicitly discussed. The reason that the cosmological constant can now 
vanish, where it could not in bosonic gravity, can be directly traced to 
the fact that in higher-derivative supergravity the ostensibly auxiliary 
field in the gravity supermultiplet becomes a new, propagating degree of 
freedom. This phenomenon was first presented in the context of $D=4$, 
$N=1$ supergravity by Ferrara, Grisaru, and van Nieuwenhuizen 
\cite{NPB-138-430}. This new degree of freedom, by extending the range of vacuum 
solutions, allows vacuum states with  zero energy. This result opens the 
door to non-trivial supergravity vacua playing a role in particle 
physics. One is led to ask whether such vacua have any demonstrable 
physical effect. The answer is a resounding yes. We will show that 
generically these non-trivial, flat spacetime supergravitational vacua 
spontaneously break supersymmetry! It seems plausible to us that, if 
this result persists in four-dimensional, $N=1$ supergravity, it 
represents a new approach to spontaneous supersymmetry breaking in 
phenomenological supergravity and, perhaps, superstring theories 
\cite{PLB-388-512}. We have recently shown that this phenomenon does, indeed, 
exist in four dimensions. This work will be presented elsewhere 
\cite{PP-UPR-685T}.

This paper is organized as follows. In Section~2, we discuss 
higher-derivative bosonic gravity in two dimensions, introducing the 
method of Legendre transformations and reducing these theories to 
canonical second-order form. In Section~3, such theories are generalized 
to quadratic $(1,1)$ supergravity. The structure of these theories is 
discussed and, using a supersymmetric generalization of the method of 
Legendre transformations, they also are reduced to a canonical 
second-order form. A similar method, within the context of $D=4$, $N=1$ 
supergravity using the compensator formalism, was given in \cite{PLB-190-86}. 
The main results of this paper are to be found in Section~4. Here we 
restrict ourselves to a specific class of models and explore their 
vacuum structure. We demonstrate explicitly that they generically 
contain non-trivial vacua with vanishing cosmological constant and that 
these vacua spontaneously break supersymmetry. We close the section by 
showing how $(1,1)$ supersymmetry allows the cosmological constant at 
non-trivial vacua to vanish. We present our conclusions and a few 
closing remarks in Section~5. Relevant details about $(1,1)$ 
supergravity \cite{JPA-12-393}, as well as the notation we will use, are 
given in a brief Appendix.

\section{Gravity in Two Dimensions}

Einstein gravity in two dimensions is trivial, since its action,
\begin{equation}
\label{EG}
S = \int d^2x\, \sqrt{-g} R,
\end{equation}
is an integral over a total divergence. This can be easily seen by using 
the fact that any two-dimensional space is conformally flat. Hence, we 
can always chose the conformal gauge in which the metric is given by
\begin{equation}
\label{confg}
g_{mn} = e^\sigma \eta_{mn}.
\end{equation}
A straightforward calculation then shows that the Lagrangian in the 
above action is just~$\nabla^2 \sigma$.

How would one go about writing non-trivial gravity theories in 
two dimensions? One way is to include a scalar field, $\lambda$. A 
simple non-trivial action for the metric $g_{mn}$ and the scalar field 
$\lambda$ can be written as
\begin{equation}
\label{L}
S = \int d^2x\, \sqrt{-g} e^\lambda R.
\end{equation}
Since the term $\sqrt{-g}e^\lambda R$ is not a total divergence, the 
equations of motion of both gravity and the scalar field $\lambda$ are 
non-trivial. We can naturally generalize action \eqref{L} by adding an 
arbitrary potential term for $\lambda$,
\begin{equation}
\label{LV}
S = \int d^2x\, \sqrt{-g} [e^\lambda R - V(\lambda)].
\end{equation}
The equations of motion for $\lambda$ and $g_{mn}$ derived from 
\eqref{LV} are
\begin{equation}
\begin{split}
\label{LVEM}
R &= e^{-\lambda} \frac{dV}{d\lambda}, \\
\nabla_m \nabla_n e^\lambda - g_{mn} \nabla^2 e^\lambda &=  \tfrac12  
g_{mn} V(\lambda).
\end{split}
\end{equation}
respectively. The physical content of this theory is most easily 
extracted by choosing the conformal gauge \eqref{confg}. In this gauge, 
the action becomes
\begin{equation}
S = \int d^2x\, [e^\lambda \nabla^2 \sigma - e^\sigma V(\lambda)].
\end{equation}
The fields $\sigma$ and $\lambda$ enter the above Lagrangian on, more or 
less, an equal footing. Expanding the integrand as a power series in 
$\lambda$ and $\sigma$ yields
\begin{equation}
\label{siglam}
S = \int d^2x\, [\nabla^2\sigma + \lambda \nabla^2\sigma - m^2\lambda^2 
+ \cdots ]
\end{equation}
where $m^2=d^2V/d\lambda^2|_{\lambda=0}$. The first term is a total 
divergence, so can be dropped. The second and third terms are quadratic 
kinetic energy and mass terms respectively. All other terms are 
higher-order interactions that we will ignore for the time being. In 
order to diagonalize the quadratic kinetic energy term in the action, we 
make the field redefinitions
\begin{equation}
\label{exp}
\phi_+ = \frac{\lambda+\sigma}{2}, \qquad \phi_-=\frac{\lambda-
\sigma}{2}.
\end{equation}
Solving for $\lambda$ and $\sigma$ in terms of $\phi_+$ and $\phi_-$, 
and substituting back into the quadratic piece of the action 
\eqref{siglam}, yields
\begin{equation}
S_Q = \int d^2x\, [-\nabla^m\phi_+ \nabla_m\phi_+ + \nabla^m\phi_- 
\nabla_m\phi_- - m^2 (\phi_+ + \phi_-)^2 ].
\end{equation}
The field $\phi_+$ has a proper kinetic energy term and, hence, is a 
physically propagating degree of freedom. However, $\phi_-$ has a 
kinetic energy term with the opposite sign. It follows that it is a 
degree of freedom with ghost-like propagating behavior. The origin of 
this ghost-like behavior is clearly the off-diagonal quadratic coupling 
inherent in the $e^\lambda R$ term. In four dimensions one can remove 
the $\lambda$--$R$ coupling by performing a conformal transformation of 
the metric. However, in two dimensions this is no longer true. Consider 
the conformal transformation
\begin{equation}
\bg_{mn} = \Omega^2 g_{mn},
\end{equation}
where $\Omega$ is an arbitrary conformal factor, to be chosen later. A 
little manipulation shows that the action \eqref{LV} takes the form
\begin{equation}
S = \int d^2x\, \sqrt{-\bg} [ e^\lambda \bR + 2 e^\lambda \bn^2 \ln 
\Omega - \Omega^{-2} V(\lambda) ].
\end{equation}
This conformally transformed action has the same $\lambda$-$R$ cross 
term. We cannot get rid of this coupling, no matter how we choose the 
conformal factor. This is in accordance with the previous statement that 
action \eqref{LV} describes two degrees of freedom. If we had succeeded 
in reducing the gravity part of the action to pure Einstein form, the 
gravity sector of the theory would be trivial and the action could 
describe only a single degree of freedom, that of the field $\lambda$. 
Be this as it may, one might still want to make a conformal 
transformation, in order to grow an explicit kinetic energy term for 
$\lambda$, and put the action in a more canonical form. Let us choose 
the conformal factor to be
\begin{equation}
\Omega = e^\lambda.
\end{equation}
The conformally transformed action then takes the form,
\begin{equation}
\label{LKV}
S = \int d^2x\, \sqrt{-\bg} [e^\lambda \bR - 2 e^\lambda (\bn\lambda)^2 
- e^{-2\lambda} V(\lambda) ].
\end{equation}
The net effect of the conformal transformation is to grow an explicit 
kinetic energy term for $\lambda$, with a sigma-model factor in front of 
it. The equations of motion for $\lambda$ and $\bg_{mn}$ obtained by 
varying action \eqref{LKV} are
\begin{equation}
\label{LKVEM}
\begin{gathered}
\bR = -2(\bn\lambda)^2 - 4 \bn^2 \lambda - 2 e^{-
3\lambda} V(\lambda) + e^{-3\lambda} \frac{dV}{d\lambda}, \\
 \bn_m\bn_n e^\lambda - \bg_{mn} \bn^2 e^\lambda - e^\lambda \bg_{mn} 
(\bn\lambda)^2 + 2 e^\lambda \bn_m\lambda \bn_n\lambda =  \tfrac12 
\bg_{mn} e^{-2\lambda} V(\lambda).
\end{gathered}
\end{equation}
respectively. Obviously, if we work in the $\bg_{mn}$ frame with action 
\eqref{LKV}, and then chose conformal gauge, we will get different 
formulae for the fields $\phi_+$ and $\phi_-$, but one of them will 
certainly be a ghost-like. This situation is unavoidable in such 
theories. It is important to note that the $\bg_{mn}$ equations of 
motion for constant field $\lambda_0$ immediately require that 
$V(\lambda_0)=0$. The $\lambda$ equation simply determines 
$\bar{R}$ given $\lambda_0$. Thus, the condition that constant 
$\lambda_0$ be a vacuum solution of the theory is given by 
$V(\lambda_0)=0$ and the value of $dV/d\lambda|_{\lambda_0}$ need not be 
specified. It is only if we further demand that the spacetime has zero 
cosmological constant, that is $\bar{R}=0$, that we must take 
$dV/d\lambda|_{\lambda_0}=0$. This is exactly the reverse of the 
situation in all dimensions greater than two, where the vacuum condition 
is that $dV/d\lambda|_{\lambda_0}=0$, whereas $V(\lambda_0)=0$ implies 
vanishing cosmological constant.

The non-trivial gravity theory that we considered, defined by action 
\eqref{LV}, seems a priori to be an unmotivated and ad hoc choice. 
However, as we will now show, this is not the case. First, let us 
consider an apparently unrelated way of getting non-trivial gravity in 
two dimensions by introducing higher-derivative gravitational terms. In 
four dimensions there are, in addition to $R$, two more tensors that 
can be used in constructing the Lagrangian, namely the Ricci tensor and 
the Riemann tensor. In two dimensions, however, both the Ricci and 
Riemann tensors can be expressed in terms of $R$. Hence, we have only 
one tensor at our disposal for constructing Lagrangians. The simple 
choice of $R$ itself as the Lagrangian was discussed above and is 
trivial. However, choosing an arbitrary function of $R$ for the 
Lagrangian yields non-trivial gravitation, as we will now show. Consider 
the action
\begin{equation}
\label{fR}
S = \int d^2x\, \sqrt{-g} f(R),
\end{equation}
where $f$ is an arbitrary real function of the scalar curvature $R$. 
The scalar curvature $R$ is a function of the metric field $g_{mn}$, 
its first-order derivatives $\partial_\ell g_{mn}$, and its second-order 
derivatives $\partial_p\partial_\ell\ g_{mn}$. Hence, the equations of 
motion for the metric field of the above action \eqref{fR} are expected 
to be fourth-order equations. Such theories are referred to as 
higher-derivative theories of gravitation. The importance of studying 
higher-derivative theories in two dimensions is not only that they 
provide, as we will show, a way of making gravity non-trivial, but that 
they also mimic many properties of higher-derivative theories of gravity 
in four dimensions, which arise in phenomenologically relevant theories 
such as supergravity models and string theory.

The equations of motion for the metric $g_{mn}$ derived from the above 
action are
\begin{equation}
f' R_{mn} - \tfrac12 f g_{mn} + g_{mn} \nabla^2 f' - \nabla_m\nabla_n f' 
= 0,
\end{equation}
which, for a generic choice of $f$, are fourth-order differential 
equations as expected. Using the fact that, in two dimensions, the Ricci 
tensor can be written as
\begin{equation}
R_{mn} = \tfrac12 g_{mn} R,
\end{equation}
we can write the equations of motion in the form
\begin{equation}
\label{fREM}
\nabla_m\nabla_n f' - g_{mn} \nabla^2 f' = \tfrac12 g_{mn} ({f'}R -f).
\end{equation}
Let us make the identification
\begin{equation}
\label{eL}
e^\lambda = f'(R),
\end{equation}
which can be inverted to give the scalar curvature $R$ in terms of the  
field $\lambda$ as
\begin{equation}
R = X(e^\lambda)
\end{equation}
where $X$ denotes the functional inverse of $f'$. Furthermore, define a 
potential energy for $\lambda$ as
\begin{equation}
\label{V}
V(\lambda) = e^\lambda X(e^\lambda) - f(X(e^\lambda)).
\end{equation}
With these identifications, equation \eqref{fREM} is in exactly the same 
form as the second equation in \eqref{LVEM}, with the potential energy 
specified in \eqref{V}. Also note that the first equation in 
\eqref{LVEM} is also satisfied as can easily be shown by differentiating 
the potential in \eqref{V}. Therefore, the higher-derivative equation of 
motion \eqref{fREM} is equivalent to the two second-order equations of 
motion \eqref{LVEM}. Of course, if we conformally transform \eqref{fREM} 
and compare it with \eqref{LKVEM}, we will be able to make the same 
identification, although we have to work a little harder.

This equivalence can be established on the level of the actions by the 
method of Legendre transformations. This elegant procedure is, in fact, 
much easier to apply. We start by introducing an auxiliary field $X$ and 
the transformed action
\begin{equation}
\label{aux}
S = \int d^2x\, \sqrt{-g} [f'(X)(R-X) + f(X)].
\end{equation}
The auxiliary field $X$ has the equation of motion
\begin{equation}
f''(X) (R-X) = 0
\end{equation}
Provided that $f''(X)\neq0$, this gives $X=R$, which when substituted 
into \eqref{aux}, gives back action \eqref{fR}. Now we can define a 
scalar field $\lambda = \ln f'(X)$, such that the action \eqref{aux} 
represents a Legendre transform from the variable $R$ to the variable 
$e^\lambda$. Writing the above action in terms of $\lambda$ we find that
\begin{equation}
S = \int d^2x\, \sqrt{-g} [e^\lambda R - V(\lambda) ],
\end{equation}
where 
\begin{equation}
V(\lambda) = e^\lambda X(e^\lambda) - f(X(e^\lambda)).
\end{equation}
Comparing this result with action \eqref{LV}, we conclude that the 
generic higher-derivative gravitation theory described by action 
\eqref{fR} is equivalent to the non-trivial gravity-plus-scalar theory 
discussed earlier. Of course, one can also perform a conformal 
transformation on the metric $g_{mn}$ to put the theory in the canonical 
form \eqref{LKV}, if one so desires.

As a concrete example of this formalism, let us consider the quadratic 
higher-derivative action
\begin{equation}
\label{R^2}
S = \int d^2x\, [ R + \epsilon R^2 ].
\end{equation}
We introduce an auxiliary field $X$ and write an equivalent action to 
\eqref{R^2} as
\begin{equation}
\label{R^2X}
S = \int d^2x\, [ (1+2\epsilon X)(R-X) + (X+\epsilon X^2)].
\end{equation}
The equation of motion of $X$ is
\begin{equation}
\label{XEM}
X = R.
\end{equation}
Substituting \eqref{XEM} into \eqref{R^2X} gives the original 
higher-derivative action \eqref{R^2}. This establishes the equivalence 
of the higher-derivative action \eqref{R^2} and the second-order action 
\eqref{R^2X}. Now define
\begin{equation}
\label{el}
e^\lambda = 1+2\epsilon X.
\end{equation}
Using this definition, action \eqref{R^2X} becomes
\begin{equation}
S = \int d^2x [ e^\lambda R - V(\lambda)],
\end{equation}
where $V(\lambda)$ is given by
\begin{equation}
\label{R2V}
V(\lambda) = \frac{1}{4\epsilon} (e^\lambda-1)^2.
\end{equation}

To conclude, by writing a higher-derivative theory of gravity, not only 
did we make gravity a non-trivial propagating degree of freedom, but we 
also introduced another propagating degree of freedom, the field 
$\lambda$. Classically the theory is completely equivalent to the 
gravity-plus-scalar theory described by the action \eqref{LV}. 
Furthermore, one of these degrees of freedom is ghost-like.

\section{Supergravity in Two Dimensions}

Supercharges are decomposable into left- and right-chiral species. In 
two dimensions, the supersymmetry algebra can have $p$ left supercharges 
and $q$ right supercharges. This is referred to as $(p,q)$ 
supersymmetry. In this paper, we will be interested in $(1,1)$ 
supersymmetry only, since this is the closest analog to the 
phenomenologically relevant $N=1$ supersymmetry in four dimensions. The 
theory of $(1,1)$ supergravity was studied by Howe \cite{JPA-12-393}, and we 
will use his results and notation. We present the relevant formulae, and 
set the notation, in the Appendix. Howe found that the supergravity 
multiplet consists of a graviton $\du{e}{m}{a}$, a gravitino 
$\du\chi{a}\alpha$ and an auxiliary field $A$. All the geometrical 
quantities in $(1,1)$ superspace, such as the curvature and the torsion, 
can be expressed in terms of these component fields. Two important 
superfields are the superdeterminant
\begin{equation}
E = e\left( 1 + \tfrac{i}{2} \theta^{\alpha} \udu\gamma{a}\alpha\beta 
\chi_{a\beta} + \bar{\theta}\theta \left[\tfrac{i}{4}A + \tfrac{1}{8} 
\epsilon^{ab} {\chi_a}^\alpha \udu\gamma5\alpha\beta 
\chi_{b\beta}\right]\right),
\end{equation}
and the real scalar superfield
\begin{equation}
S = A + i \theta^{\alpha}\Sigma_{\alpha}+\tfrac{i}{2} 
\bar{\theta}\theta C,
\end{equation}
where
\begin{equation}
\begin{split}
C &= -R-\tfrac12 {\chi_a}^\alpha \udu\gamma{a}\alpha\beta 
\psi_{\beta} + \tfrac{i}{4} \epsilon^{ab} {\chi_a}^{\alpha} 
\udu\gamma5\alpha\beta \chi_{b\beta} A-\tfrac12A^{2}, \\
\Sigma_{\alpha} &= -2 \epsilon^{ab} \udu\gamma5\alpha\beta 
{\mathscript{D}}_{a}\chi_{b \beta} - \tfrac12 \udu\gamma{a}\alpha\beta 
\chi_{a\beta}A.
\end{split}
\end{equation}
If $S$ vanishes, so does the curvature and the superspace is flat. 
Einstein supergravity in two dimensions is given by 
\begin{equation}
\label{ESG}
S = 2i \int d^2x d^2\theta E S.
\end{equation}
If we expand this action in components, we find that there is no 
contribution from the gravitino $\du\chi{a}\alpha$, while the bosonic 
part is just Einstein gravity \eqref{EG}. Therefore, minimal 
supergravity in two dimensions is also trivial, with no propagating 
degrees of freedom.

How would one go about writing non-trivial supergravity theories in 
two dimensions? We can try to supersymmetrize the non-trivial gravity 
theories of the previous section. The supersymmetrization of gravity 
coupled to a scalar field of the form \eqref{LV} was discussed by 
various authors \cite{PLB-258-97,PRD-47-1569}. In this paper, we are more 
interested in the 
supersymmetric analog of action \eqref{fR}; that is, in 
higher-derivative supergravitation.

A naive supersymmetrization would be to consider a Lagrangian that is a  
general function of the superfield $S$. However, since the spacetime 
curvature scalar $R$ occurs as the highest component of $S$, any 
function $f(S)$, prior to the elimination of auxiliary fields, will 
contain only $R$ rather than arbitrary powers of $R$. The equation of 
motion of the $A$ field is algebraic, which means $A$ is an auxiliary 
field. Eliminating $A$ from the Lagrangian, we find, generically, that 
$A$ is expressed in terms of fractional powers of $R$. Inserting the 
solution of the algebraic equation of motion of $A$ into the Lagrangian 
does lead to higher powers of $R$. However, for all but the simplest 
functions $f(S)$, the equation of motion of $A$ cannot be solved in 
closed form. For this reason, we will not consider such theories in this 
paper. A simpler supersymmetrization is to consider Lagrangians that 
contain higher powers of $R$ prior to any elimination of auxiliary 
fields.

With this in mind, we would like to construct a superfield that has $R$ 
is its lowest component. We can form such a superfield by taking the 
$\theta$-derivative of $S$ twice. Since we need a covariant object, we 
have to consider $\D^\alpha \D_\alpha S$. This superfield, along with 
$S$ itself, can be used to construct superfields with arbitrary higher 
powers of the spacetime curvature $R$. However, it can be easily shown 
that any quadratic or higher power of $\D^\alpha \D_\alpha S$ leads to 
equations of motion that are fourth-order in the field $A$. A 
fourth-order equation of motion of a scalar field contains a ghost-like 
degree of freedom \cite{PR-79-145}. For this reason, we demand that the 
Lagrangian does not contain any powers of $\D^\alpha \D_\alpha S$ higher 
than unity. Are there any other covariant terms that can be used to 
construct a higher-derivative supergravity Lagrangian? Any terms with 
higher derivatives with respect to $\theta$, for example $\D^\alpha 
\D_\alpha \D^\beta \D_\beta S$, or spacetime derivatives, such as 
$\nabla^m \nabla_m S$, lead to $\nabla^m \nabla_m R$ in the component 
field action. If we demand that the supersymmetric theory be at most 
fourth-order, then terms like $\D^\alpha \D_\alpha \D^\beta \D_\beta S$ 
and  $\nabla^m \nabla_m S$ are excluded.

According to the above discussion, the general fourth-order supergravity 
action is given by
\begin{equation}
\label{fgS}
S = 2i \int d^2x d^2\theta\,  E [f(S) + i g(S) \D^\alpha S \D_\alpha S],
\end{equation}
where $f$ and $g$ are two arbitrary real functions of the superfield 
$S$. Note that we have made an integration by parts in writing the 
Lagrangian in the above form. What is the dynamical content of action 
\eqref{fgS}? If we expand the above action in components, using the 
expressions introduced in the Appendix, and perform the $\theta$ 
integrals, we obtain a Lagrangian of the following form
\begin{equation}
\mathscript{L} = \mathscript{L}_{\text{Boson}} + \mathscript{L}_{\text{Fermion}} 
+ \mathscript{L}_{\text{Boson-Fermion}},
\end{equation}
where
\begin{equation}
\label{LB}
\mathscript{L}_{\text{Boson}} = e \left[ - (f'(A) - 2 g(A) A^2 ) R -2 g(A) 
R^2 + 2 g(A) (\nabla A)^2 - \tfrac12 ( f'(A) A^2 + g(A) A^4) \right].
\end{equation}
The $\mathscript{L}_{\text{Fermion}}$ and 
$\mathscript{L}_{\text{Boson-Fermion}}$ terms are rather complicated 
expressions and will not be needed here. The explicit appearance of 
$R^2$ confirms that Lagrangian \eqref{fgS} is indeed a 
supersymmetrization of $R^2$ gravity. Note that no higher-powers of 
$R$ appear. Thus despite including two arbitrary functions $f$ and $g$, 
we only have a supersymmetric extension of $R^2$ rather than of any 
arbitrary power of $R$. The $\D^\alpha S \D_\alpha S$ term not only has 
the effect of introducing $R^2$ directly in the Lagrangian, but also of 
growing a kinetic energy term for the scalar field $A$, as can be seen 
from \eqref{LB}. What does this mean? It means that the field $A$, is no 
longer auxiliary. It is now a propagating field. How many degrees of 
freedom are described by the bosonic part of the action given in 
\eqref{LB}? As we have seen in the previous section, $R + R^2$ gravity 
describes two real degrees of freedom. Hence, \eqref{LB} describes three 
real degrees of freedom, two coming from the higher derivative $R^2$ 
term, and the third being the once auxiliary field $A$.

Since the action \eqref{fgS} is supersymmetric, one should expect 
supersymmetric partners for these three degrees of freedom. There should 
be three fermions propagating along with these three bosonic fields. 
Indeed, a direct computation of the fermionic part of the Lagrangian 
shows that the equation of motion of the gravitino $\du\chi{a}\alpha$ is 
third-order. A single fermionic degree of freedom is described by a 
first-order differential equation. Only one initial condition is 
required to solve the Cauchy problem of such a field. A higher-order 
differential equation implies the existence of more degrees of freedom. 
In particular, a third-order differential equation describes three 
fermionic degrees of freedom, since three initial conditions are 
required to solve the Cauchy problem. These are the three fermionic 
degrees of freedom associated with the three bosonic degrees of freedom.

To better understand the dynamical content of \eqref{fgS}, we would like 
to transform it into a second-order theory, in much the same way as we 
transformed the bosonic $R+R^2$ theory in \eqref{R^2} into a 
second-order theory in the previous section. Hence, we are tempted to 
introduce a superfield in much the same way we introduced the scalar 
field $\lambda$ in the previous section. But, as discussed before, 
$R+R^2$ bosonic gravity describes only two degrees of freedom, one of 
which is the graviton. Therefore a single real field $\lambda$ is all 
that is required to describe the extra degree of freedom and to 
transform the theory to second-order form. Here, as we have shown, there 
are three degrees of freedom, one of which is the graviton. Thus we need 
two superfields to describe the extra degrees of freedom and to 
transform the theory into a second-order form. We will denote these real 
superfield by $\Phi$ and $\Lambda$. Apart from this subtlety, there is 
not much difference between the method of reduction as applied to 
bosonic theories or supersymmetric theories. Action \eqref{fgS} is 
equivalent to the second-order action
\begin{align}
S &= 2i\int d^2xd^2\theta E [f(\Phi)+ i g(\Phi)\D^\alpha\Phi 
\D_\alpha\Phi + e^\Lambda(S-\Phi)] \notag \\
&= 2i\int d^2xd^2\theta E [e^\Lambda S + i g(\Phi)\D^\alpha\Phi 
\D_\alpha\Phi + f(\Phi) - e^\Lambda\Phi ].
\label{fgPL}
\end{align}
The superfield equations of motion of $\Lambda$ and $\Phi$ are given by
\begin{equation}
\label{PL}
\begin{split}
\Phi &= S, \\
e^\Lambda &= f'(S) - i g'(S) \D^\alpha S \D_\alpha S - 2 i g(S) 
\D^\alpha \D_\alpha S,
\end{split}
\end{equation}
respectively. Here $\Lambda$ acts as a Lagrange multiplier, with the 
effect of setting $\Phi$ equal to $S$. Substituting $\Phi=S$ into 
\eqref{fgPL} gives us back the original action \eqref{fgS}. Action 
\eqref{fgPL} is the supersymmetric extension of the bosonic action 
\eqref{LV}. The superfield $\Phi$ is a propagating superfield with an 
explicit kinetic energy term. The superfield $\Lambda$ is also a 
propagating superfield, due to its coupling with the superfield $S$. It 
is clear that quadratic supergravity is equivalent to non-trivial 
supergravity (due to the $e^\Lambda$ factor in front of $S$) coupled to 
two new scalar superfield degrees of freedom.

What exactly are the new propagating degrees of freedom in terms of the 
original variables $\du{e}m{a}$, $\du\chi{a}\alpha$, and $A$? Gravity 
itself is propagating, so the graviton $\du{e}m{a}$ and the gravitino 
$\du\chi{a}\alpha$ are propagating degrees of freedom. Besides this, the 
equations of motions of $\Phi$ and $\Lambda$ give the new degrees of 
freedom in terms of the original variables. Consider $\Phi$ first. It 
follows from \eqref{PL} and \eqref{S} that
\begin{equation}
\begin{split}
\label{Phi}
\Phi| &= A, \\
\Phi|_\theta &= -2 \epsilon^{ab} \udu\gamma5\alpha\beta 
{\mathscript{D}}_{a}\chi_{b \beta} - \tfrac12 \udu\gamma{a}\alpha\beta 
\chi_{a\beta}A.
\end{split}
\end{equation}
Therefore, $A$ is a propagating degree of freedom, as we have already 
seen. Its fermionic superpartner is a complicated function of the first  
derivative of the gravitino and $A$ itself. Secondly, consider 
$\Lambda$. The lowest component of $\Lambda$ is set equal to the natural 
logarithm of the lowest component of the second line in \eqref{PL}, 
which is a particular, but involved, function of both $A$ and $R$. The 
supersymmetric partner of this field, namely, the fermionic field in 
$\Lambda$, is even more involved as a function of the original fields. 
However, at the linearized level, the components of $\Lambda$ can be 
evaluated. They are given by
\begin{equation}
\begin{split}
\Lambda| &= \ln f'(0) +\frac{f''(0)}{f'(0)}A+\frac{2g(0)}{f'(0)} R, \\
\Lambda|_\theta &= - \frac{2 f''(0)}{f'(0)} \e^{ab} 
\udu\gamma5\alpha\beta D_a \chi_{b\beta} - \frac{8 g(0)}{f'(0)} 
\udu\gamma{a}\alpha\beta \e^{bc} \udu\gamma5\beta\gamma D_a D_c 
\chi_{b\gamma}.
\end{split}
\end{equation}
The $\Lambda|$ equation is the analog of expression \eqref{eL} in the 
pure bosonic case. There we saw that it is roughly $R$ that is 
propagating. Here we see that it is a mixture of both $R$ and $A$ that 
make up the second propagating scalar field. The fermionic degree of 
freedom, $\Lambda|_\theta$, contains the first- as well as the 
second-derivative of the gravitino. These complicated expressions for 
the propagating degrees of freedom show how powerful the method of 
Legendre transformations really is. 

As stated above, both $\Phi$ and $\Lambda$ are propagating superfields. 
This can be made more explicit if we put the above action \eqref{fgPL} 
in canonical form by performing a super-Weyl transformation to grow an 
explicit kinetic energy term for $\Lambda$. Consider the super-Weyl 
transformation discussed in the Appendix,
\begin{equation}
\label{super-Weyl}
\begin{split}
\bar{E} &= e^\Lambda E, \\
\bar{S} &= e^{-\Lambda} S + ie^{-\Lambda} \D^\alpha\D_\alpha 
\Lambda.
\end{split}
\end{equation}
Under such a transformation, action \eqref{fgPL} becomes 
\begin{equation}
\label{sot}
S=2i\int d^2xd^2\theta \bar{E} [e^\Lambda\bar{S} + i 
e^\Lambda\bar{\D}^\alpha \Lambda \bar{\D}_\alpha\Lambda + i 
g(\Phi) \bar{\D}^\alpha\Phi \bar{\D}_\alpha\Phi + e^{-\Lambda} 
f(\Phi) - \Phi].
\end{equation}
This is the supersymmetric extension of the bosonic action \eqref{LKV}. 

In the following, we will drop the bar for notational simplicity. The 
Superfields $\Phi$ and $\Lambda$ can be expanded into component fields 
as
\begin{equation}
\label{LP}
\begin{split}
\Phi &= \phi + i\theta^\alpha\pi_\alpha + \tfrac{i}{2} 
\bar{\theta}\theta F, \\
\Lambda &= \lambda + i\theta^\alpha\xi_\alpha + \tfrac{i}{2} 
\bar{\theta}\theta G.
\end{split}
\end{equation}
Inserting these expressions, as well as the expansions of $E$ and $S$, 
into \eqref{sot} gives a component field Lagrangian in which the fields 
$A,F$, and $G$ are auxiliary. The equations of motion of these fields 
are
\begin{equation}
\label{AuxOS}
\begin{split}
A &= 2 e^{-\lambda} \phi - e^{-2\lambda} f(\phi) + i \xi^\alpha 
\xi_\alpha, \\
F &= \frac{1}{4g(\phi)} \left[ -1 + e^{-\lambda} f(\phi) - i g'(\phi) 
\pi^\alpha \pi_\alpha \right], \\
G &= \tfrac12 e^{-\lambda} \phi - \tfrac12 e^{-2\lambda} f(\phi) + 
\tfrac{i}{2} \xi^\alpha \xi_\alpha.
\end{split}
\end{equation}
Substituting these expressions for $A,F$, and $G$ into the Lagrangian 
gives the component field Lagrangian for the propagating graviton 
$\du{e}{m}{a}$, gravitino $\du\chi{a}\alpha$, scalar fields $\phi$ and 
$\lambda$, and their fermionic partners $\pi$ and $\xi$. We get a 
Lagrangian of the form
\begin{equation}
\mathscript{L} = \mathscript{L}_{\text{Boson}} + \mathscript{L}_{\text{Fermion}} 
+ \mathscript{L}_{\text{Boson-Fermion}}.
\end{equation}
The bosonic part of the Lagrangian is
\begin{equation}
\label{fgLB}
\mathscript{L}_{\text{Bosnon}} = e \left[ e^\lambda R - 2 e^\lambda 
(\nabla\lambda)^2 - 2 g(\phi) (\nabla\phi)^2 - V(\phi,\lambda) \right],
\end{equation}
where the potential $V$ is given by
\begin{equation}
\label{NPB-138-430}
V(\phi,\lambda) = \frac{1}{8g(\phi)} \left[ 1 - 2 e^{-\lambda} ( 
f'(\phi) + 2  \phi^2 g(\phi) ) + e^{-2\lambda} (f'(\phi)^2 + \phi 
f(\phi) g(\phi) ) \right].
\end{equation}
The $\mathscript{L}_{\text{Fermion}}$ and $\mathscript{L}_{\text{Boson-
Fermion}}$ parts of the Lagrangian are rather complicated and will not 
be presented here. Ignoring the fermionic fields, the bosonic equations 
of motion can be deduced from \eqref{fgLB} and \eqref{NPB-138-430}. They are 
given by
\begin{equation}
\begin{gathered}
R = -2 (\nabla\lambda)^2 - 4 \nabla^2 \lambda + e^{-\lambda} 
\frac{\partial}{\partial\lambda}V(\phi,\lambda) \\
2 g'(\phi) (\nabla\phi)^2 + 4 g(\phi) \nabla^2 \phi = 
\frac{\partial}{\partial\phi} V(\phi,\lambda) \\
\nabla_m\nabla_n e^\lambda - g_{mn} \nabla^2 e^\lambda - e^\lambda 
g_{mn}(\nabla\lambda)^2+2e^\lambda \nabla_m\lambda\nabla_n\lambda \\
 -g(\phi) g_{mn} (\nabla\phi)^2 + 2 g(\phi) \nabla_m\phi \nabla_n\phi = 
\tfrac12 g_{mn} e^{-2\lambda} V (\phi,\lambda)
\end{gathered}
\end{equation}
for fields $\lambda$, $\phi$ and $g_{mn}$ respectively. For constant 
fields $\phi_0$ and $\lambda_0$, these equations reduce to
\begin{equation}
\label{consVacua}
\begin{split}
\left. \frac{\partial V}{\partial\lambda} \right|_{(\phi_0,\lambda_0)} 
&= e^{\lambda_0} R \\
\left.\frac{\partial V}{\partial\phi}\right|_{(\phi_0,\lambda_0)}&=0\\
V(\phi_0,\lambda_0) &=0 
\end{split}
\end{equation}
It follows that the equations specifying the constant vacua of the 
theory are $\partial V/\partial\phi|_{\phi_0}=0$ and 
$V(\phi_0,\lambda_0)=0$, whereas, in general, $\partial 
V/\partial\lambda|_{\lambda_0}$ is arbitrary. It is only if we demand 
that the vacuum has vanishing cosmological constant, that is $R=0$, 
that $\partial V/\partial\lambda|_{\lambda_0}=0$. We emphasize again 
that these conditions are different than the associated conditions for 
theories in higher-dimensional spacetimes. We will investigate these 
vacua in the next section. 

\section{Non-Trivial Vacua and Supersymmetry Breaking}

In this section, we would like to simplify the problem of finding the 
vacua of the higher-derivative supergravity by considering a concrete 
example for the functions $f$ and $g$. Having made such a choice, we 
will evaluate the scalar potential $V(\phi,\lambda)$ and look for a 
constant vacuum state $\phi=\phi_0$ and $\lambda=\lambda_0$ with zero 
cosmological constant. At such a minimum, one has to make sure that $g$ 
is non-negative to avoid having a ghost superfield $\Phi$. The simple 
choice of $g(S)$ to be a positive constant suffices in this regard. 
Furthermore, one sees from \eqref{NPB-138-430} that linear or higher-order terms 
in $g$ yield a rational potential $V$ as a function in $\phi$, making 
the problem less tractable. For these reasons, we will chose
\begin{equation}
\label{g}
g(S)=c,
\end{equation}
where $c$ is a real positive constant. For simplicity, we will chose $f$ 
to be a general cubic polynomial in $S$ with real coefficients
\begin{equation}
\label{f}
f(S) = a + S + bS^2 + dS^3.
\end{equation}
The coefficient of the linear term $S$ can be chosen to be unity by 
adjusting the overall normalization of the action. 

With these choices, the potential energy \eqref{NPB-138-430} becomes
\begin{align}
V &= \frac{1}{8c}\left\{ 1-2 \left[ 1+2b \phi 
+\left(2c+3d\right)\phi^{2}\right] e^{-\lambda} + \left[1 + 
4\left(b+ac\right) \phi + 2\left(2b^{2} + 2c+3d\right)\phi^{2} 
\right.\right. \notag \\
& \hspace*{0.15in} \left.\left. + 4b \left(c+3d\right) 
\phi^{3}+d\left(4c+9d\right) \phi^{4}\right] e^{-2\lambda} \right\}.
\end{align}
We now solve generically for constant vacua of this theory with 
vanishing cosmological constant. It follows from \eqref{consVacua} that 
we must solve the equations
\begin{equation}
\begin{split}
\diff{V}{\lambda} &= 0, \\
\diff{V}{\phi} &=0, \\
V(\phi,\lambda) &= 0.
\end{split}
\end{equation}
For any values of the parameters $a,b,c,d$ there exists an extremum at 
$\phi_0=\lambda_0=0$. The potential $V$ vanishes at this point, without 
the need to adjust any of the parameters. We will refer to this point as 
the trivial extremum. We are interested to see if other, non-trivial, 
extrema with vanishing potential $V$ exist. We find that there is 
precisely one non-trivial extremum given by
\begin{equation}
\label{vacuum}
\begin{split}
\phi_0 &= \frac{-2b}{3\left(c+2d\right)}, \\
1-e^{-\lambda_0} &= \frac{-4b^{2} \left(c+3d\right) }{9c^{2}-
4b^{2}\left(c+3d\right) +36d\left(c+d\right)}, \\
a &= \frac{4}{27}\frac{b^{3}}{\left(c+2d\right)^{2}},
\end{split}
\end{equation}
for every choice of the parameters $b,c$, and $d$. The condition on $a$ 
is required to ensure that the potential $V$ is zero. We will, 
henceforth, restrict our discussion to the class of theories satisfying 
this condition on $a$. Any theory in this class is parametrized by 
$b,c$, and $d$. When $b\neq0$, the theory has two distinct extrema. 
However, when $b=0$ the two extrema become identical and the theory has 
only the trivial extremum at $\phi_0=\lambda_0=0$. For different values 
of parameters $b,c,d$ the non-trivial extremum \eqref{vacuum} can be a 
local maximum, a saddle point, or a minimum. We will come back to this 
point later. 

The supersymmetry transformation laws for the fermions, given in 
\eqref{SUSYS} and \eqref{SUSYPhi} in the Appendix, are
\begin{equation}
\begin{split}
\delta \chi_{m\alpha} &= 2 (\partial_m \tau_\alpha + \tfrac12 \omega_m 
\udu\gamma5\alpha\beta \tau_\beta ) + \tfrac12 \du \gamma{m\alpha}\beta A 
\tau_\beta, \\
\delta \pi_\alpha &= [\udu\gamma{m}\alpha\beta (\partial_m\phi 
-\tfrac{i}{2} \du\chi{m}\gamma \pi_\gamma )] \tau_\beta-F\tau_\alpha, \\
\delta \xi_\alpha &= [\udu\gamma{m}\alpha\beta (\partial_m\lambda 
-\tfrac{i}{2} \du\chi{m}\gamma \xi_\gamma )]\tau_\beta-G\tau_\alpha, \\
\end{split}
\end{equation}
where $\tau_\alpha$ is the supersymmetry transformation parameter. Note 
that if any of the auxiliary fields $A,F,G$ develops a non-vanishing 
vacuum expectation value, the corresponding fermion will develop an 
inhomogeneous piece in its supersymmetry transformation law, which 
signals supersymmetry breaking. Using \eqref{AuxOS}, \eqref{g}, and 
\eqref{f}, we find that the auxiliary fields all vanish at the trivial 
extremum $\phi_0=\lambda_0=0$. Hence, supersymmetry is never broken at 
that point. On the other hand, the auxiliary fields evaluated at the 
extremum \eqref{vacuum} are given by
\begin{equation}
\begin{split}
A_0 &= - \frac{18b(24d^3+36d^2c-8b^2d^2-4b^2dc+18dc^2+3c^3)}{P^2}, \\
F_0 &= - \frac{2b^2}{P}, \\
G_0 &= - \frac{12b^3c(c+2d)}{P^2},
\end{split}
\end{equation}
where $P$ is a polynomial in $b,c$, and $d$ given by
\begin{equation}
\label{P}
P=c\left(9c-4b^{2}+36d\right)-12d\left(b^{2}-3d\right).
\end{equation}
Note that apart from the case $b=0$, which makes the extremum 
\eqref{vacuum} coincide with trivial extremum $\phi_0=\lambda_0=0$, at 
least one of these vacuum expectation values of the auxiliary fields, 
namely $F_0$, is non-zero. We will assume from now on that $b\neq0$. 
With such a choice, we can conclude that supersymmetry is broken at the 
non-trivial extremum \eqref{vacuum}.

Since supersymmetry is broken, there should exist a massless fermion in 
the theory, a goldstino. In order to see this, we need to consider the 
fermion mass-matrix. The fermionic part of the Lagrangian, 
$\mathscript{L}_{\text{Fermion}}$, splits into three terms, the kinetic 
energy term, the mass term, which is quadratic in the fermions, and a 
four-fermi interaction term. The fermionic mass term, evaluated at the 
non-trivial extremum \eqref{vacuum}, is given by
\begin{align}
\label{MF}
\mathscript{L}_{\text{M-Fermion}} & =  e \left( 
m_{11}i\pi^{\alpha}\pi_{\alpha}+m_{22} i \zeta^{\alpha} 
\zeta_{\alpha}+m_{33} \epsilon^{ab} {\chi_a}^{\alpha} 
\udu\gamma5\alpha\beta \chi_{b\beta} + 
2m_{12}i\pi^{\alpha}\zeta_{\alpha} \right. \notag \\
& \hspace*{0.4in} \left. + 2m_{13}i\pi^{\alpha}\udu\gamma{a}\alpha\beta 
\chi_{a\beta} + 2m_{23}i\zeta^{\alpha}\udu\gamma{a}\alpha\beta 
\chi_{a\beta} \right),
\end{align}
where
\begin{equation}
\begin{split}
m_{11} &= \frac{9bc\left(c+2d\right)}{P}, \\
m_{22} &= \frac{-2b\left(-2b^{2}+3c+6d\right)}{P}, \\
m_{33} &= \frac{2b^{3}c}{3\left(c+2d\right)P}, \\
m_{12} &= \frac{-3c\left(3c-4b^{2}+12d\right)+12d\left(b^{2}-
3d\right)}{2P}, \\
m_{13} &= \frac{-2b^{2}c}{P}, \\
m_{23} &= \frac{bc\left(9c-8b^{2}+36d\right)-12bd\left(b^{2}-3d\right)} 
{6\left(c+2d\right)P},
\end{split}
\end{equation}
and $P$ is the polynomial defined in \eqref{P}. We can diagonalize the 
fermion mass matrix as follows. First define
\begin{equation}
\tilde{\chi}_{a\alpha} = \chi_{a\alpha} + 2 \du\gamma{a\alpha}\beta 
(m_{13} \pi_\beta + m_{23} \xi_\beta).
\end{equation}
Then
\begin{align}
m_{33} i \epsilon^{ab} \du{\tilde{\chi}}a\alpha \udu\gamma5\alpha\beta 
\tilde{\chi}_{b\beta} = & m_{33} i \epsilon^{ab} \du\chi{a}\alpha 
\udu\gamma5\alpha\beta \chi_{b\beta} + 
2m_{13}i\pi^{\alpha}\udu\gamma{a}\alpha\beta \chi_{a\beta} + 
2m_{23}i\zeta^{\alpha}\udu\gamma{a}\alpha\beta \chi_{a\beta} \notag \\
& + 2 \frac{m_{13}^2}{m_{33}} i \pi^\alpha \pi_\alpha + 2 
\frac{m_{23}^2}{m_{33}} i \xi^\alpha \xi_\alpha + 4 
\frac{m_{13}m_{23}}{m_{33}} i \pi^\alpha \xi_\alpha.
\label{cs}
\end{align}
Note that we have used the assumption that $b\neq0$, which implies 
supersymmetry is broken, since otherwise $m_{33}$ would be zero and the 
above computation would break down. However, keeping this assumption in 
mind, we can proceed and substitute \eqref{cs} into \eqref{MF}. The 
fermion mass term, then, takes the form 
\begin{equation}
\Lag_{\text{M-Fermion}} = e\left( m_{33} i \epsilon^{ab} 
\du{\tilde{\chi}}a\alpha \udu\gamma5\alpha\beta \tilde{\chi}_{b\beta} + 
\tilde{m}_{11} i \pi^\alpha\pi_\alpha + \tilde{m}_{22} i 
\xi^\alpha\xi_\alpha + 2 \tilde{m}_{12} i \pi^\alpha\xi_\alpha \right),
\end{equation}
where
\begin{equation}
\begin{split}
\tilde{m}_{11} &= m_{11} - 2 \frac{m_{13}^2}{m_{33}}, \\
\tilde{m}_{22} &= m_{22} - 2 \frac{m_{23}^2}{m_{33}}, \\
\tilde{m}_{12} &= m_{12} - 4 \frac{m_{13}m_{23}}{m_{33}}.
\end{split}
\end{equation}
Since the ``shifted'' gravitino $\du{\tilde{\chi}}{a}\alpha$ is a 
mixture of the original fields $\du\chi\mu\alpha,\pi^\alpha$, and 
$\xi^\alpha$, its supersymmetry transformation law changes accordingly. 
The auxiliary field part of its transformation, evaluated at the 
non-trivial extremum \eqref{vacuum}, is now given by
\begin{equation}
\delta \tilde{\chi}_{a\alpha} = \du\gamma{a\alpha}\beta 
\left( \tfrac12 A_0 + m_{13} F_0 + m_{23} G_0 \right) \tau_\beta = 0.
\end{equation}
In other words, we find that the shifted gravitino transforms 
homogeneously. We now diagonalize the $2\times2$ mass matrix for the 
fermions $\pi_\alpha$ and $\xi_\alpha$. We find that it has the two 
eigenfields
\begin{equation}
\begin{split}
\tilde{\pi}_\alpha &= \pi_\alpha + h \xi_\alpha, \\
\tilde{\xi}_\alpha &= \xi_\alpha - h \pi_\alpha,
\end{split}
\end{equation}
where $h$ is given by
\begin{equation}
h = - \frac{P}{6bc(c+2d)}.
\end{equation}
The fermion $\tilde{\pi}$ has a non-vanishing mass given by 
$\tilde{m}_{11}$, whereas the mass of fermion $\tilde{\xi}$ vanishes. 
The auxiliary-field piece in the supersymmetry transformation law of the 
new field $\tilde{\pi}$ is given by
\begin{equation}
\delta \tilde{\pi}_\alpha = - (F_0 + h G_0) \tau_\alpha 
=0,
\end{equation}
which means that $\tilde{\pi}$ transforms homogeneously. On the other 
hand, $\tilde{\xi}$ transforms with an auxiliary field piece given by
\begin{align}
\delta \tilde{\xi}_\alpha & = -( G_0 - h F_0) \tau_\alpha 
\notag \\
&= - ( 1 + h^2) G_0 \tau_\alpha \neq 0.
\end{align}
That is, $\tilde{\xi}$ transforms inhomogeneously. The diagonalization 
of the fermion mass term is now complete. The diagonal form can be 
written as
\begin{equation}
\Lag_{\text{M-Fermion}} = e\left( m_{33} i \epsilon^{ab} 
\du{\tilde{\chi}}a\alpha \udu\gamma5\alpha\beta \tilde{\chi}_{b\beta} + 
\tilde{m}_{11} i \tilde{\pi}^\alpha \tilde{\pi}_\alpha \right).
\end{equation}
The vanishing mass for $\tilde{\xi}$ implies that $\tilde{\xi}$ is a 
Goldstone fermion, in accordance with the spontaneously broken 
supersymmetry at this extremum. This conclusion is further strengthened 
by the fact that the gravitino, $\du{\tilde{\chi}}{a}\alpha$, has 
acquired a non-vanishing mass. The only field which transform 
inhomogeneously is the massless Goldstone fermion $\tilde\xi^\alpha$.

The above analysis is true at the non-trivial extremum \eqref{vacuum}, 
no matter whether it is a local maximum, saddle point, or minimum. 
However, we are specifically interested in the stable vacuum of the 
theory. The fact that any non-trivial theory of gravitation in two 
dimensions contains a ghost-like degree of freedom, as discussed in 
Section~2, makes it unclear as to exactly what one means by stable 
vacuum state. However, for the theory under investigation, if we turn 
gravity off, we still have two physical degrees of freedom. In this 
case, we can require that the theory be stable at an extremum of the 
potential in the usual sense; that is, any fluctuation of the fields 
around the extremum should increase the energy. This is true if and only 
if the extremum locally minimizes the potential. We find that the 
non-trivial extremum \eqref{vacuum} is a local minimum of the potential 
when the two conditions 
\begin{equation}
\begin{split}
2d + c &< 0, \\
48 d^3 + 68 d^2 c + 32 d c^2 + 5 c^3 &< 0,
\end{split}
\end{equation}
are simultaneously satisfied. For example, the choice $b=c=-d=1$, which 
satisfy the above two conditions, makes the $2\times 2$ Hessian scalar 
mass matrix positive definite, and, hence, ensures that \eqref{vacuum} 
is a local minimum. It follows that there exists a class of theories in 
which there is a non-trivial stable vacuum state with zero cosmological 
constant which breaks the $(1,1)$ supersymmetry spontaneously. This 
result will obviously continue to hold for more general choices of the 
functions $f(S)$ and $g(S)$.

In a previous paper, we considered the most general theory of quadratic 
bosonic gravitation in four dimensions \cite{PRD-53-5583}. We showed that the 
only stable vacuum in this theory is the trivial vacuum with vanishing 
cosmological constant. In a subsequent paper, we generalized our results 
to higher-derivative bosonic gravitation beyond the quadratic level 
\cite{PRD-53-5597}. We showed that such theories still possess a trivial 
vacuum with vanishing cosmological constant but, unlike the quadratic 
case, they generically have non-trivial vacua as well. These vacua, 
however, are always characterized by a non-vanishing cosmological 
constant. Hence, any non-trivial vacua in these bosonic theories must be 
deSitter or anti-deSitter spaces, typically with a radius of curvature 
of the order of the inverse Planck mass. For this reason, such theories 
are of little interest to particle physics. We find that exactly 
analogous behavior occurs in the bosonic two-dimensional 
higher-derivative gravity theories discussed in Section~2 of this paper. 
Let us digress briefly to make these bosonic two-dimensional properties 
explicit, before returning to the supergravitational case. It turns out 
to be more convenient to discuss these properties using the original 
metric $g_{mn}$ rather than the conformally transformed metric 
$\bar{g}_{mn}$. Therefore, we will use metric $g_{mn}$ for the 
remainder of this section. Consider the quadratic bosonic action 
\eqref{R^2}. We reduce this theory to second-order form by introducing a 
single scalar field $\lambda$ with the potential energy given by 
\eqref{R2V}. Combining \eqref{XEM} and \eqref{el}, we find that the 
relation between $\lambda$ and $R$ is
\begin{equation}
\label{eLR2}
e^\lambda = 1 + 2 \e R.
\end{equation}
If we demand that spacetime has vanishing cosmological constant, or, 
what is the same thing, $R=0$, then we must have $\lambda_0=0$. In 
order for this to be a vacuum of the theory, it follows from 
\eqref{LVEM} that the potential energy must satisfy $V(0)=0$. Using the 
potential \eqref{R2V}, one can easily verify that this is the case. 
Moreover, $\lambda_0=0$ is the only zero of potential \eqref{R2V}. 
Hence, there are no other vacua, whether they correspond to flat 
spacetime or not. This is exactly equivalent to the bosonic 
four-dimensional quadratic gravity case. Now consider the more general 
higher-derivative bosonic gravitation of action \eqref{fR}. This theory, 
once again, is reduced to second-order form by the introduction of a 
single scalar field $\lambda$. The relation between $R$ and $\lambda$ 
is now given by the more complicated equation \eqref{eL}, namely
\begin{equation}
\label{lmeom}
e^\lambda = f'(R).
\end{equation}
Although this relation is more general than the quadratic case 
\eqref{eLR2}, the fact remains that the zero cosmological constant 
condition, $R=0$, corresponds to a unique solution at $\lambda_0=\ln 
f'(0)$. This can, without loss of generality, always be normalized to 
$\lambda_0=0$. As before, it follows from \eqref{LVEM} that this point 
is a vacuum of the theory only if the potential energy satisfies 
$V(0)=0$, which is equivalent to the condition $f(0)=0$. Thus, provided 
there is no constant piece in the function $f$, we find that the theory 
has a trivial vacuum state at $\lambda_0=0$. Generally, the potential 
function is considerably more complicated than the simple potential 
\eqref{R2V} in the quadratic case. Unlike the quadratic potential 
\eqref{R2V}, it generically has more than one zero. However, from the 
condition \eqref{lmeom}, we see that any non-trivial vacuum different 
from $\lambda_0=0$ will not correspond to flat spacetime. Instead, it 
will have a constant non-vanishing curvature $R$ and correspond to 
deSitter or anti-deSitter spacetime with non-zero cosmological 
constant. Again, this is in direct analogy with the associated 
four-dimensional case.

Given these results for bosonic gravitation, it appears all the more 
remarkable that, in this paper, we have shown that two-dimensional 
higher-derivative supergravitational theories allow non-trivial vacua 
with vanishing cosmological constant. In fact, as we have discussed, 
even quadratic supergravity theories possess non-trivial vacua 
corresponding to $R=0$. It follows that such theories are far more 
relevant to particle physics. It is worth, therefore, a discussion of 
why supersymmetric theories differ from bosonic ones in this crucial 
issue.

Let us consider the case of quadratic supersymmetric gravitation given 
by \eqref{fgS}. The reduction of this theory to second-order form 
\eqref{fgPL} requires the introduction of two supermultiplets with 
scalar degrees of freedom $\phi$ and $\lambda$. In order to deduce the 
relation between $\phi$, $\lambda$, and $R$, we expand \eqref{fgPL} 
into component form and compute the equations of motion for $F$, $G$, 
and $\lambda$. Eliminating the auxiliary fields from the $\lambda$ 
equation gives the desired relation
\begin{equation}
R = \frac{1}{4g(\phi)} \left(e^\lambda - f'(\phi) - \tfrac12 \phi^2 
g(\phi)\right).
\end{equation}
This relation is the supersymmetric analog of \eqref{eLR2}. It should be 
clear that requiring $R$ to vanish no longer singles out a unique 
choice of $\phi$ and $\lambda$. Indeed, there may be many values of 
$\phi$ and $\lambda$ that are compatible with $R=0$. In the particular 
case we studied, where $g$ and $f$ are given by \eqref{g} and \eqref{f} 
respectively, the above relation becomes
\begin{equation}
R=\frac{1}{4c}\left(e^\lambda-1-2b\phi-(3d+2c)\phi^2\right).
\end{equation}
It can be easily verified that both the trivial fields 
$\phi_0=\lambda_0=0$ and the non-trivial fields \eqref{vacuum} are 
compatible with $R=0$. Furthermore, both the trivial and non-trivial 
solutions satisfy, by construction, the conditions 
$V(\phi_0,\lambda_0)=0$ and $\partial V/\partial 
\phi|_{(\phi_0,\lambda_0)}=0$ and, hence, are vacua of the theory. It is 
clear that it is the new degree of freedom $\phi$, which is introduced 
by supersymmetry, that allows non-trivial vacuum solutions corresponding 
to zero cosmological constant to exist. As shown in \eqref{Phi}, $\phi$ 
is directly related to the ``auxiliary'' field $A$ of the $(1,1)$ 
supergravity multiplet that propagates, and becomes physical, in 
higher-derivative theories. The possibility of non-trivial vacuum 
solutions with zero cosmological constant makes higher-derivative 
supergravitation much more relevant to particle physics than the bosonic 
theories previously discussed.

\section{Conclusions}

We have constructed the most general quadratic $(1,1)$ supergravitation 
theory in two dimensions. We have shown that this theory is reducible to 
a second-order form by the introduction of two real scalar 
supermultiplets. We have evaluated the scalar potential for the 
second-order theory and presented an explicit class of examples which 
possess a non-trivial stable vacuum state with zero cosmological 
constant that spontaneously breaks the $(1,1)$ supersymmetry. This 
result is quite general and leads to the main conclusion of this paper: 
two-dimensional $(1,1)$ supergravity theories generically possess 
stable, flat spacetime, but non-trivial, vacua that spontaneously break 
supersymmetry. Supersymmetry breakdown is due to non-trivial vacuum 
expectation values for the extra scalar degrees of freedom that arise 
directly from the super-zweibein in higher-derivative theories. That is, 
supersymmetry is broken by supergravity itself.

In our opinion, this result represents a new approach to the theory of 
spontaneous supersymmetry breaking. If it could be extended to 
four-dimensional $N=1$ supergravity, this new method of supersymmetry 
breaking would have obvious applications to phenomenological 
supersymmetric theories. With this in mind, we have recently shown that, 
indeed, exactly the same phenomenon occurs in $D=4$, $N=1$ quadratic 
supergravitation \cite{PP-UPR-685T}. It follows that higher-derivative 
supergravity might serve as a natural mechanism for spontaneously 
breaking supersymmetry in phenomenologically interesting particle 
physics models. The results of our ongoing investigations will be 
presented elsewhere.

\renewcommand{\theequation}{A.\arabic{equation}}

\section*{Appendix: Two-Dimensional (1,1) Superspace}

The structure of $(1,1)$ superspace was studied by Howe \cite{JPA-12-393}. 
The $(1,1)$ superspace has coordinates $z^M=(x^m,\theta^\mu)$, where $m$ 
and $\mu$ can both take on two values. We will use ($m,n,\ldots$) for 
spacetime indices, ($a,b,\ldots$) for tangent space indices, and 
($\alpha,\beta,\ldots$) for spinor indices. The bosonic metric and the 
anti-symmetric tensor are
\begin{equation}
\begin{split}
\eta_{ab}&=\text{diag}(-1,+1), \qquad \epsilon_{ab}=-\epsilon_{ba}, 
\qquad \epsilon_{01}=1, \\
\epsilon^{ab}&=-\epsilon_{ab}, \qquad \e_{ab} \e^{bc} = \du\delta{a}c.
\end{split}
\end{equation}
The fermionic anti-symmetric ``metric'' is given by
\begin{equation}
\begin{split}
\epsilon_{\alpha\beta} &= \epsilon^{\alpha\beta}, \qquad \e_{12}=1=-
\e_{21},
\qquad \e_{11}=\e_{22}=0, \\
\e_{\alpha\beta} \e^{\beta\gamma} &= - \du\delta\alpha\gamma, \qquad
\e_{\alpha\beta} \e^{\alpha\beta} = 2.
\end{split}
\end{equation}
The $\gamma$-matrices are chosen to be real, satisfying
\begin{equation}
\udu\gamma{a}\alpha\beta \udu\gamma{b}\beta\gamma = \eta^{ab}
\du\delta\alpha\gamma - \e^{ab}\udu\gamma5\alpha\gamma.
\end{equation}
where $\gamma^5=\gamma^0\gamma^1$. This implies the relations
\begin{equation}
[\gamma^a,\gamma^b] = -2\e^{ab}\gamma^5, \qquad
[\gamma^a,\gamma^5]=2\e^{ab}\gamma_b, \qquad 
\gamma^a\gamma^5=\e^{ab}\gamma_b.
\end{equation}
An explicit representation is given by
\begin{equation}
\udu\gamma{0}\alpha\beta = \begin{pmatrix} 0 & 1 \cr -1 & 0 
\end{pmatrix}, \qquad
\udu\gamma1\alpha\beta = \begin{pmatrix} 0 & 1 \cr  1 & 0  
\end{pmatrix}, \qquad
\udu\gamma5\alpha\beta = \begin{pmatrix} 1 & 0 \cr  0 & -1 
\end{pmatrix}.
\end{equation}

The geometry of the $(1,1)$ superspace is determined by the 
super-zweibein $\du{E}MA$ and the connection $\du\Omega{B}A$. There are 
two important two-forms, the torsion and curvature defined by
\begin{equation}
\begin{split}
T^A &= \D E^A = \tfrac12 E^C\wedge E^B \du{T}{BC}A, \\
\du{R}AB &= d\du\Omega{A}B + \du\Omega{A}C \wedge \du\Omega{C}B = 
\tfrac12 E^D
\wedge E^C\du{R}{CD,A}B.
\end{split}
\end{equation}
They satisfy the Bianchi identities,
\begin{equation}
\begin{split}
\D T^A &= E^B \wedge \du{R}BA, \\
\D \du{R}AB &= 0. 
\end{split}
\end{equation}
Howe imposed the following set of constraint on the supertorsion
\begin{equation}
\du T{\beta\gamma}a = 2i \ud\gamma{a}{\beta\gamma}, \qquad 
\du{T}{\beta\gamma}\alpha =
\du{T}{bc}a =0.
\end{equation}
He found that all the components of the torsion and curvature can then 
be written in terms of a single superfield $S$. If $S$ vanishes, so does 
the curvature and the space is flat. In Wess-Zumino gauge, every tensor 
can be expressed in terms of only three component fields, namely, the 
zweibein $\du{e}ma$, the Rarita-Schwinger field $\du\chi{m}\alpha$, and 
an ``auxiliary'' scalar field $A$. The supervolume element $E$ is given 
by
\begin{equation}
E=e\left( 1 + \tfrac{i}{2} \theta^{\alpha} \udu\gamma{a}\alpha\beta 
\chi_{a\beta} + \bar{\theta}\theta \left[\tfrac{i}{4}A + \tfrac{1}{8} 
\epsilon^{ab} {\chi_a}^\alpha \udu\gamma5\alpha\beta 
\chi_{b\beta}\right]\right),
\end{equation}
where $\bar{\theta}\theta = \theta^\alpha \theta_\alpha$. The 
superfield $S$ is given by
\begin{equation}
\label{S}
S=A + i \theta^{\alpha}\Sigma_{\alpha}+\tfrac{i}{2} 
\bar{\theta}\theta C,
\end{equation}
where
\begin{equation}
\begin{split}
C &= -{R}-\tfrac12 {\chi_a}^\alpha \udu\gamma{a}\alpha\beta 
\psi_{\beta} + \tfrac{i}{4} \epsilon^{ab} {\chi_a}^{\alpha} 
\udu\gamma5\alpha\beta \chi_{b\beta} A-\tfrac12A^{2}, \\
\Sigma_{\alpha} &= -2 \epsilon^{ab} \udu\gamma5\alpha\beta 
{\mathscript{D}}_{a}\chi_{b \beta} - \tfrac12 \udu\gamma{a}\alpha\beta 
\chi_{a\beta}A.
\end{split}
\end{equation}

Howe showed that the generalization of Weyl transformations to 
superspace, compatible with the above constraints, is given by the 
super-Weyl transformations
\begin{equation}
\begin{split}
{{{\bar{E}}}_{M}}^{a} & = \Lambda \du{E}Ma, \\
\du{\bar{E}}M\alpha & = \Lambda^{1/2} \du{E}M\alpha 
- \tfrac{i}{2} \Lambda^{-1/2} \du{E}Ma \du\gamma{a}{\alpha\beta} 
\D_\beta \Lambda, \\
\du{\bar{E}}aM & = \Lambda^{-1} \du{E}aM + i \Lambda^{-2} 
\du\gamma{a}{\alpha\beta} \D_\beta \Lambda \du{E}\alpha{M}, \\
\du{\bar{E}}\alpha M & = \Lambda^{-1/2} \du{E}\alpha M. \\
\end{split}
\end{equation}
One can compute the change in the superfield $S$ under these 
transformations, finding
\begin{equation}
\bar{S}=\Lambda^{-1} S + i \Lambda^{-3} \D_\alpha\Lambda 
\D^\alpha\Lambda
- i \Lambda^{-2} \D_\alpha\D^\alpha \Lambda.
\end{equation}
Howe also showed that every $(1,1)$ superspace is super-conformally 
flat.

We will consider theories of supergravity coupled to matter. Matter 
superfields $\Phi$ are real scalar superfields having an expansion of 
the form
\begin{equation}
\Phi = \phi + i \theta^\alpha \pi_\alpha 
+ \tfrac{i}{2} \bar{\theta}\theta F.
\end{equation}
For these fields we have a choice for their super-Weyl weight. We will 
choose zero super-Weyl weight for all the matter fields considered in 
this paper.

Howe also deduced the supersymmetry transformations for the 
gravitational multiplet,
\begin{equation}
\label{SUSYS}
\begin{split}
\delta \du{e}m{a} & = i \tau^\alpha \udu\gamma{a}\alpha\beta 
\chi_{m\beta}, \\
\delta \chi_{m\alpha} &= 2 (\partial_m \tau_\alpha + \tfrac12 \omega_m 
\udu\gamma5\alpha\beta \tau_\beta ) + \tfrac12 \du \gamma{m\alpha}\beta A 
\tau_\beta, \\
\delta A &= i \tau^\alpha \psi_\alpha.
\end{split}
\end{equation}
where $\tau_\alpha$ is the gauge parameter and $\omega_m$ is the spin 
connection. We will also need the supersymmetry transformations of a 
matter superfield $\Phi$, which are given by
\begin{equation}
\label{SUSYPhi}
\begin{split}
\delta \phi &= i \tau^\alpha \pi_\alpha, \\
\delta \pi_\alpha &= [\udu\gamma{m}\alpha\beta (\partial_m\phi 
- \tfrac{i}{2} \du\chi{m}\gamma \pi_\gamma )]\tau_\beta-F\tau_\alpha, \\
\delta F &= i \tau^\alpha \udu\gamma{m}\alpha\beta \left[ -(\partial_m 
\pi_\beta + \tfrac12 \omega_m \udu\gamma5\beta\gamma  \pi_\gamma )
+ \tfrac12 \udu\gamma{n}\beta\gamma (\partial_n\phi 
- \tfrac{i}{2}\du\chi{n}\delta\pi_\delta) \chi_{m\gamma} 
- \tfrac12 F \chi_{m\beta} \right].
\end{split}
\end{equation}

\section*{Acknowledgments}

This work was supported in part by DOE Grant No.\ DE-FG02-95ER40893 and 
NATO Grand No.\ CRG-940784.

\end{document}